\newcommand{\bq}{\begin{equation}} \newcommand{\eq}{\end{equation}}
\newcommand{\ba}{\begin{eqnarray}}
\newcommand{\ea}{\end{eqnarray}}

\newcommand{\nl}{\\ \nonumber}

\def\ifmath#1{\relax\ifmmode #1\else $#1$\fi}%
\let\ifmathx0
\def\fixmath{\def\ifmath{\noexpand\ifmathx}}%

\def\ovMZ{\ifmath{{\overline M}_Z}}
\def\MZ{\ifmath{M_Z}}

\def\ovMZ{\ifmath{{\overline m}_Z}}
\def\ovGZ{\ifmath{{\overline \Gamma}_Z}}

\def\gam{\ifmath{\gamma}}%
\def\Zo{\ifmath{\mathrm {Z}}}

\def\Rgf{\ifmath{\mathrm {R}^f_\gam}}

\def\RZfi{\ifmath{\mathrm {R_Z}^{fi}}}

\def\Ffin{\ifmath{\mathrm {F}_n^{fi}}}

\documentstyle[12pt,epsf,epsfig]{article}
\begin{document}
\pagestyle{empty}
 \begin{flushleft}
hep-ph/9712435
 \end{flushleft}
\begin{center}
{\huge 
Study of the $Z$ Boson at LEP
 \vspace*{0.5cm}               
}


{\large 
Tord Riemann
}
\vspace*{0.0cm}
 
\begin{normalsize}
{\it
Deutsches Elektronen-Synchrotron DESY, Institut f\"ur Hochenergiephysik
\\ 
IfH Zeuthen, Platanenallee 6, D-15738 Zeuthen, Germany
}
\end{normalsize}
\end{center}
 
 \vspace*{.0cm} 

\begin{abstract}
The $Z$ line shape is measured at LEP with an accuracy at the
per mille level.
Usually it is described in the Standard Model of electroweak
interactions with account of quantum corrections.
Alternatively, one may attempt different 
model-independent approaches in order to extract quantities like mass
and width of the $Z$ boson.
If a fit deviates from that in the standard approach, this may give hints for
New Physics contributions.
I describe two model-independent approaches and compare their
applications to LEP data with the Standard Model approach.
\end{abstract}


  
\section{Introduction}
From 1989 till 1995 about 18 millions of $Z$ bosons have been produced at LEP1
and about 200\,000 at SLC. 
Due to this, and due to the lack of direct hints for the existence of a Higgs
boson, the $Z$ boson and its interactions became for several years the
central theme of tests of the Standard Model
\cite{Glashow:1961,Weinberg:1967,Salam:1968rm}, 
recently accompanied by the discovery of the $t$ quark at
the Tevatron \cite{Abe:1995hr,Abachi:1995iq}.

The predictions of the electroweak Standard Model depend on the
particle masses, fermion mixings, and one coupling constant.
One of the best measured parameters is by now the $Z$ boson mass.

In 1983, at the $p{\bar p}$ collider SPS (CERN) 
the $Z$ boson 
was discovered \cite{Arnison:1983mk,Bagnaia:1983zx} and the mass
could be determined 
at that time with an accuracy of several GeV; in 1986:
\ba
M_Z = 92.6 \pm
1.7 ~ {\mathrm GeV}
\ea

At the end of 1989 LEP1 and SLC started operation and dominated the precision
experiments for tests of the electroweak Standard Model for a decade.
This may be exemplified by quoting the following improvements of precision from
August 1989 \cite{Altarelli:1989gg} till October 1997 \cite{Altarelli:1997sk}:
\ba
1989 ~~~~~~ &\to& ~~~~~~ 1997
\nonumber \\
\nl
M_Z = 91.120 \pm 0.160 ~ {\mathrm GeV} ~~&\to&~~ 91.186\,7 \pm 0.002\,0   
~ {\mathrm GeV}  
\\
\sin^2\theta_w^{eff} = 0.233\,00 \pm 0.002\,30 ~~&\to&~~ 0.231\,52 \pm
0.000\,23 
\\
{m}_{t}^{pred} = 130 \pm 50  ~ {\mathrm GeV} ~~&\to&~~ m_t^{meas} =
175.6 \pm 5.5 
~ {\mathrm GeV}  
\\
M_H^{pred} >  ~ {\mathrm few~GeV} ~~&\to&~~ \geq 77  ~ {\mathrm GeV}
\\
\alpha_s(M_Z) = 0.110 \pm 0.010 ~~&\to&~~ 0.119 \pm 0.003
\ea

\section{The $Z$ line shape}
The $Z$ boson may be studied as a resonance at LEP from a measurement
of the cross-section 
\ba
e^+ e^- \to (\gamma, Z) \to {\bar f}f ( + n \gamma)
\ea
as a function of the beam energy; see Fig.~1.
The determinations of mass $M_Z$ and width $\Gamma_Z$ 
are dominated by hadron production in a small region around the peak:
$|\sqrt{s} - M_Z | < 3$ GeV.
The $Z$ is not a pure Breit-Wigner resonance.
We want to study a $2 \to 2$ process with intermediate $Z$, but have
also virtual photon exchange.
In addition, there are huge $2 \to 3$ contributions due to initial state
radiation (ISR) of photons and due to final state radiation (FSR).
Further, many virtual corrections are contributing as quantum
  corrections: 
vertex insertions,
self energy insertions,
box diagrams, and all their iterations.

\begin{figure}[thbp]
 \vspace*{-5.0cm}
 \hspace*{-3.0mm}
\mbox{
       \epsfig{file=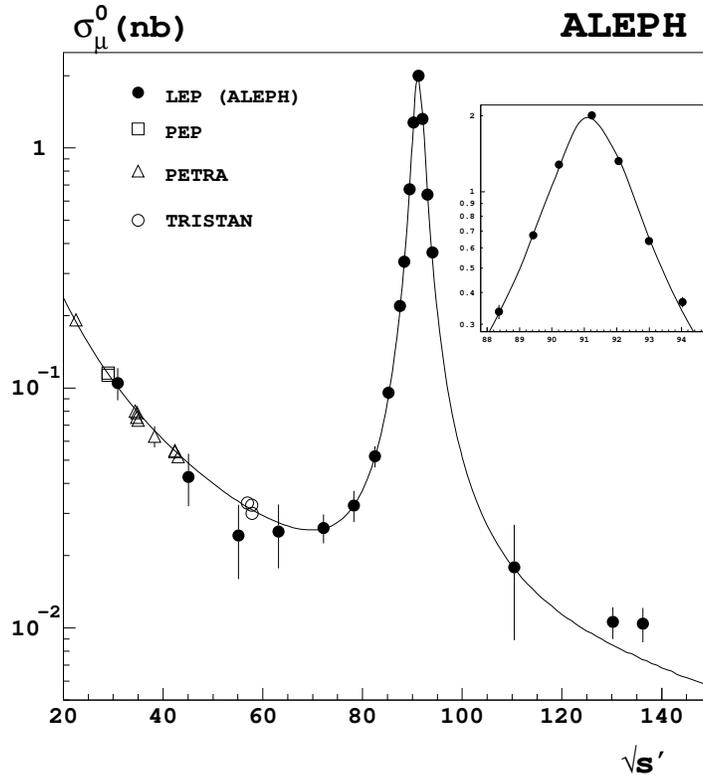,height=20cm,%
                   clip=%
}}
 \vspace*{-5.6cm}
\begin{center}
\caption{\it
The muon production cross-section over a wide energy range; figure by
courtesy of Frederic Teubert. 
\label{sigma}
}
\end{center}
\end{figure}

\section{Real photonic corrections
\label{qed}
}
The QED corrections may be taken into account in a universal way by
the following convolution formula
(\cite{Bardin:1989cw,Bardin:1991de,Bardin:1995aa,PChristova:1997aa} 
and references therein):
\ba
\sigma(s)
&=&\int\frac{ds'}{s}\sigma_0(s')\, \rho\left(\frac{s'}{s}\right)
~+~\int\frac{ds'}{s}\sigma_0^{int}(s,s')
\, \rho^{int}
\left(\frac{s'}{s}\right)
\label{sigqed2}
\ea
\begin{itemize}
\item
$\rho(s'/s)$ -- the radiator 
describes initial and final state radiation,
including leading higher order effects and soft photon exponentiation;
\item
$\sigma_0(s')$ -- the basic scattering cross-section, which is the
object of investigation.
\end{itemize}
The $\rho^{int}(s'/s)$ takes into account the 
initial-final state interference effects which
are comparatively small (a few per mille) near the $Z$ resonance 
but are bigger off the resonance, and
$\sigma_0^{int}(s,s')$ is a function similar to $\sigma_0(s')$, but suppressed
if $\rho^{int}(s'/s)$ is small.

\section{Method (I):  A model-independent ansatz
\label{I}
}
QED corrections are treated by the convolution formula introduced in
section \ref{qed}.
For a careful discussion of their influence on height and location of
the $Z$ peak see \cite{Beenakker:1990ec}.
 
The following ansatz for $\sigma_0(s)$ is a good choice
without explicit reference to the Standard Model
\cite{Borrelli:1990bd,Jegerlehner:1991ed,Stuart:1991xk,Leike:1991pq}:
\ba
\sigma_0(s) =
\frac{4}{3} \pi \alpha_{em}^2
\left[ \frac{r^{\gamma}}{s} +
\frac {s\cdot R + (s - M_Z^2)\cdot J} 
{\left|s-M_Z^2 + i s \Gamma_Z/M_Z\right|^2}
\right]
\label{sigqed3}
\ea
The line shape is described by five parameters:
\begin{itemize}
\item $r^{\gamma} \sim \alpha_{em}^2(M_Z^2)$ -- may be assumed to be known
\item  $M_Z,~~ \Gamma_Z$
\item $R$ -- measure of the $Z$ peak height; related
  to $\sigma_0^{had}, \sigma_0^{lept}$ 
\item $J$ -- measure of the $\gamma Z$ interference; often fixed to
  standard model value
\end{itemize}

\subsection{$Z$ line shape fit (I)}
With the model-independent ansatz, the following nearly uncorrelated
observables may be determined from the $Z$ peak data
\cite{DRWard:1997AA,Clare:1997BB}: 
\ba
M_Z &=&  91.186\,7 \pm 0.002\,0~~ \mbox{GeV}~~~~(\delta=0.0025~\%)
\\
\Gamma_Z &=& 2.494\,8 \pm 0.002\,5~~ \mbox{GeV}~~~~(\delta=1.3~\%)
\\
\sigma_0^{had} &=& 41.486 \pm 0.053~~{\mbox{nb}}~~~~(\delta=1.9~\%)
\\
R_l = \frac{\sigma_0^{had}}{\sigma_0^{lept}} &=& 20.775 \pm
0.027~~~~(\delta=1.5~\%) 
\\
A_{FB,0}^{lept} &=& 0.017\,1 \pm 0.001\,0
\ea
Here, $M_Z, \Gamma_Z, \sigma_0^{had}$ are from $\sigma^{had}(s)$,
while $R_l$ and $A_{FB}$ from $\sigma^{lep}(s)$:
with $\sigma_0^{had(lep)}$ as hadronic (leptonic) peak cross-section, and
$A_{FB,0}^{lept}$ as forward-backward asymmetry at the peak.
These parameters are considered to be primary parameters in contrast
to derived ones, e.g. the effective
leptonic weak neutral current couplings of leptons or the effective
weak mixing angle  \cite{DRWard:1997AA,Clare:1997BB}: 
\ba
v_l &=& - 0.037\,93 \pm 0.000\,58 
\\
a_l &=& - 0.501\,03 \pm 0.000\,31
\\
\sin^{2}\theta_{w}^{eff} 
\equiv \frac{1}{4}\left( 1-\frac{v_l}{a_l}\right)
&=& 0.231\,52 \pm 0.000\,23
\label{sw1}
\ea
\section{Method (II): Virtual corrections in the Standard Model}
All virtual corrections may be written  in some theory,
e.g. the Standard Model, for massless particle production in the
following way (see
e.g. \cite{Bardin:1980fe,Bardin:1982sv} and 
\cite{Bardin:1989di,Bardin:1989tq,Bardin:1992jc1,Bardin:1992jc2,Bardin:1995a3}
and references therein): 
\ba 
\nonumber
{\bf\cal M}_{net} &\sim&
\frac{\alpha_{em}}{s} \Biggl\{
\frac{\alpha_{em}(s)}{\alpha_{em}}\left|Q_e Q_f \right| \gamma_{\beta} \otimes
\gamma_{\beta}  
+ \chi(s) \varrho_{ef} 
\Bigl[ L_{\beta} \otimes L_{\beta} 
\nl && -~ 4 s_w^2 |Q_e| \kappa_e \gamma_{\beta} \otimes L_{\beta} 
- 4 s_w^2|Q_f|  \kappa_b L_{\beta} \otimes \gamma_{\beta}
\\ && +~
16 s_w^4 |Q_eQ_f| \kappa_{eb} \gamma_{\beta} \otimes \gamma_{\beta}
\Bigr] 
\Bigr\}
\label{net}
\ea
We use short notations: $L_{\beta} = \gamma_{\beta}(1+\gamma_5 )$, 
$A_{\beta} \otimes B_{\beta} = [ \bar v_e A_{\beta} u_e ]
                        \cdot [ \bar u_b B_{\beta} v_b ]$, 
and
\ba
\chi = 
\chi(s) =  \frac{G_{\mu}}{\sqrt{2}} \frac{M_Z^2}{8\pi\alpha_{em}}
\frac{s}{s-M_Z^2+iM_Z\Gamma_Z(s)} 
, ~~
\Gamma_Z(s) = \frac{s}{M_Z^2} \Gamma_Z 
\label{basic_defs}
\ea
The effective Born cross-section now is uniquely determined once the
net matrix element ${\bf\cal M}_{net}$ is known:
\ba
\sigma_0(s)
&=&
N_c^f \sqrt{1-4m_f^2/s} ~~\frac{4\pi\alpha_{em}^2}{3s} \times
\nl
&& \Biggl\{
\Bigl(1+\frac{2 m_f^2}{s}\Bigr) 
\Bigl[
|Q_eQ_f|^2 \frac{|\alpha_{em}(s)|^2}{\alpha_{em}^2} 
+ 2 |Q_eQ_f| \Re e \Bigl( \chi  
\frac{\alpha_{em}^*(s)}{\alpha_{em}} \varrho_{ef} v_{ef} \Bigr)
\nl && +~
|\chi\varrho_{ef}|^2  (1 + |v_e|^2 + |v_f|^2+ |v_{ef}|^2 
)
\Bigr] 
-  \frac{6m_f^2}{s} |\chi\varrho_{ef}|^2 (1 + |v_e|^2)
 \Biggr\}
\label{sigeff0}
\ea
with
\ba
v_{i} &=& 1 - 4 s^2_w |Q_i|\kappa_i, ~~~~i=e,f
\\
v_{ef} &=& 1 - 4 s^2_w |Q_e| \kappa_e 
- 4 s^2_w |Q_f| \kappa_f + 16 s^4_w |Q_eQ_f| \kappa_{ef}
\ea
Further, $N_c^f=1, 3$ is the colour factor and QCD corrections also
have to be taken into account. 

The virtual corrections with higher order parts for the form factors
may be found in \cite{Bardin:1989di,Bardin:1995a3,Bardin:1998AB} and
references therein.
Further, we need expressions for $M_W$, $\Gamma_Z$, $\alpha_{em}$ and
a reasonable treatment of QCD corrections and explicit expressions for the
structures shown above.

The $W$ boson mass is:
\ba
M_{W}=M_{Z}\sqrt{1
-\sqrt{1-\frac{4\pi\alpha_{em}}{\sqrt{2}G_{\mu}M_Z^2[1-\Delta r]}}} 
\label{wmass}
\ea
  
For $\Delta r$, the $Z$ width 
\cite{Akhundov:1986fc,Jegerlehner:1986vs,Beenakker:1988pv,Bernabeu:1988me}, 
$\alpha_{em}$ 
\cite{Eidelman:1995ny,Jegerlehner:1996ab}, as well as QCD corrections
\cite{Chetyrkin:1994js3}, and all the other expressions
left out here I have to refer to
literature quoted above and to references therein.  
\subsection{$Z$ line shape fit (II)}
I quoted all the above formulae in order to demonstrate explicitly how involved
a Standard Model fit ansatz is.
The input quantities are: $\alpha_{em}, G_{\mu}$ (for $M_W$), $M_Z, m_f, M_H,
\alpha_s$. 
Some of them are precisely known (e.g. $G_{\mu}$), others are subject of
determination at LEP (e.g. $M_Z$), others are completely unknown ($M_H$).  
The $t$ quark mass may be determined from weak loop corrections at LEP or
directly from $t$ quark production at Fermilab.

Quantities like the $Z$ width or the weak mixing angle are not a subject of
fits since they are considered to be secondary quantities.
In this respect, there is a basic difference to the approach of the foregoing
section. 

The most recent Standard Model fit is \cite{DRWard:1997AA,Clare:1997BB}.
The $t$ quark mass from Fermilab is:
\ba
m_t &=& 175.6 \pm 5.5~~ \mathrm{GeV}
\ea
The global fit to all data yields:
\ba
M_Z  &=& 91.186\,7 \pm 0.002\,0 ~~  \mathrm{GeV}
\\
m_t &=& 173.1 \pm 5.4  ~~\mathrm{GeV}
\\
M_H &=& 115 ^{+116}_{-66} ~~\mathrm{GeV}
\\
\alpha_s(M_Z) &=& 0.120 \pm 0.003
\ea
In a next step, one may calculate the other quantities like the $Z$ width and
relate to values from model-independent fits. 
Whatever one does, there is no unique hint to New Physics. 
For a detailed discussion of this see \cite{Altarelli:1997sk}. 
\section{Method (III): The S-matrix approach
}
When the $Z$ boson is treated as a resonance, the S-matrix
approach may be used for its description.
This was proposed in the context of LEP physics in \cite{Stuart:1991xk}, where
the perturbation expansion in the Standard Model was studied.
In \cite{Leike:1991pq} it was proposed to use this approach for a direct fit to
LEP data and the first S-matrix fit was performed therein.
The first fit by a LEP collaboration was due to L3
\cite{Adriani:1993gk,Adriani:1993ca}. 
The treatment of asymmetries near the peak was discussed in
\cite{Leike:1991pq}. 
For the role of QED corrections to asymmetries see
\cite{Riemann:1992gv}. 

A recent survey on the definition of $Z$ mass and width and their treatment in
fermion pair production is \cite{Riemann:1997tj}.
Here, I give a short introduction to the technical essentials.
 
Consider 
four independent helicity amplitudes in the case of massless fermions:
\ba
\label{eqn:mat0}
{\cal M}^{fi}(s) = \frac{\Rgf}{s} + \frac{\RZfi}{s-s_Z} +
\sum_{n=0}^\infty \frac{\Ffin}{\ovMZ^2} \left(\frac{s-s_Z}{\ovMZ}\right)^n 
,~~\  i=1,\ldots,4.
\ea
The position of the \Zo\ pole in the complex $s$ plane is given by 
$s_{\Zo}$:
\ba
\label{szb}
s_Z = \ovMZ^2 - i \ovMZ \ovGZ.
\ea
The \Rgf\ and \RZfi\ are complex residua of the photon and the \Zo\ boson,
respectively. One may approximate (\ref{eqn:mat0}) by setting $\Ffin \to 0$.
There are four residua \RZfi\ for
$e^-_Le^+_R \to f^-_L f^+_R, 
e^-_Le^+_R \to f^-_R f^+_L,
e^-_Re^+_L \to f^-_R f^+_L, 
e^-_Re^+_L \to f^-_L f^+_R$.
The amplitudes
${\cal M}^{fi}(s)$ give rise to four cross-sections $\sigma_i$:
$\sigma_{T}^0(s),
\sigma_{FB}^0(s),
\sigma_{pol}^0(s),
\sigma_{lr}^0(s)$.
Here, the $\sigma_{T}^0$ -- the total cross-section, 
$\sigma_{FB}^0$ -- numerator of the forward-backward asymmetry, 
$\sigma_{pol}^0$ -- that of the final state polarization etc.
All these cross-sections may be parameterized by the following master
formula ($A=T,FB, \ldots$): 
\ba
\renewcommand{\arraystretch}{2.2}
\label{eqn:smxs}
\sigma_A^0(s)
&=&
\displaystyle{\frac{4}{3} \pi \alpha_{em}^2
\left[ \frac{r^{\gamma f}_A}{s} +
\frac {s r^f_A + (s - \ovMZ^2) j^f_A} {(s-\ovMZ^2)^2 + \ovMZ^2 \ovGZ^2}
\right]} + \ldots
\renewcommand{\arraystretch}{1.}
\ea
Without QED corrections, asymmetries are:
\ba
\label{eqn:mi_asy}
{\cal A}_A^0(s) = \frac{\sigma_A^0(s)}{\sigma_{T}^0(s)}
=
A_0^A + A_1^A \left(\frac{s}{\ovMZ^2} - 1 \right) + \ldots,~~~ A \neq T
\ea
They take the above extremely simple approximate form around the \Zo\
resonance. 
At LEP1, the higher order terms in the Taylor expansion may be neglected since
$(s/\ovMZ^2-1)^2  < 2 \times 10^{-4}$.
The coefficients have a quite simple form:
\ba
\label{eqn:a0}
A_0^A 
= \frac{r_A^f} {r_T^f}, ~~~~
A_1^A =
\left[ \frac{j^f_A}{r_A^f} -  \frac{j_T^f}{r_T^f} \right] A_0^A
\ea
A comment is necessary concerning the definition of mass and width of the
$Z$ boson.
The so-called pole definition with a constant width (\ref{szb})
as a natural consequence of
the S-matrix ansatz leads to different numerical values compared to
the usual Standard Model approach (\ref{basic_defs})
\cite{Bardin:1988xt,Berends:1988bg,Argyres:1995ym}.
A very precise approximation is:
\ba
\label{mmggzz}
\ovMZ & = & [ 1 + (\Gamma_Z/M_Z)^2 ]^{-\frac{1}{2}} \MZ
\approx 
M_Z - \frac{1}{2} \Gamma_Z^2/M_Z = M_Z -34~{\mbox{MeV}}
\label{eqn:mbar}
\ea
\subsection{$Z$ line shape fit (III)}
The interest of the community in an S-matrix based fit to the LEP data has
several origins.
One is the wish for a model-independent description of the resonance. 
Closely related is the question on the number of independent parameters needed
to describe the peak:
four (per channel) suffice to describe a cross-section: $M_Z, \Gamma_Z, r_T,
j_T$, provided we assume QED interactions to be understood. Among these
parameters, $Z$ mass and width are universal for all channels.
Any asymmetry introduces two additional degrees of freedom (per
channel): $r_A, j_A$. 

There are practical aspects of all this.
If the number of different energy points needed for a scan of the $Z$ peak is
asked for, the answer is at least five (four plus one) for cross-sections, 
at least three (two plus one) for asymmetries.
Further, the $\gamma Z$ interferences $j_A$ form separate  degrees of freedom. 
The $j_T$ and $M_Z$ are highly correlated. 
This became more important recently when the highest
statistics were taken, and also with the data collected at energies farer
away from the peak. 
There the interference becomes more influential.           

Recent experimental studies are summarized in \cite{Clare:1997AA}.
The data of table~\ref{table1} are obtained from the $Z$ line
shape scans at LEP which were performed mainly in 1993 and 
1995 (from table~7 of~\cite{Clare:1997AA}\footnote{Note that the table
  shows values of the complex pole mass $\ovMZ$.
The Standard Model fits use the on shell mass $M_Z$; the relation of both is  
given in (\ref{eqn:mbar}).   
}).
The biggest error correlations are shown in table~2 (from table~8
of~\cite{Clare:1997AA}). 
Including into the analysis cross-sections measured at TRISTAN energies
does not improve substantially e.g. the resolution of $M_Z$ and $j_T$
\cite{Clare:1997AA}.

\begin{table}[htbp]\centering
\begin{tabular}{|c|r@{$\pm$}l|r@{.}l|}
\hline
Parameter & \multicolumn{2}{c|}{S-matrix fit} & 
            \multicolumn{2}{c|}{SM Prediction}
\\
\hline
\hline
\ovMZ 
[GeV]      &  91.153\,4  & 0.003\,3  & \multicolumn{2}{c|}{--} 
\\
$\Gamma_Z$ [GeV] &   2.492\,4  & 0.002\,6  &  2&493\,2
\\
\hline
$r_T^{had}$      &   2.962\,3  & 0.006\,7  & 2&960\,3 
\\
$j_T^{had}$     &    0.15      & 0.15      &  0&22
\\
\hline
$r_T^{lept}$     &   0.142\,39 & 0.000\,34 &  0&142\,53
\\
$j_T^{lept}$     &   0.009     & 0.012  &  0&004
\\
\hline
$r_{FB}^{lept}$  &   0.003\,04 & 0.000\,18 &  0&002\,66
\\
$j_{FB}^{lept}$  &   0.789   & 0.013       &  0&799
\\  \hline
\end{tabular}

\vspace*{5mm}
\caption
{\it
Results from a combined LEP1 line shape fit
\label{table1}
}
\end{table}

\begin{table}[thbp]\centering
\begin{tabular}{|r@{--}l|r@{.}l|}
\hline
\multicolumn{2}{|c|}{Correlation} & \multicolumn{2}{c|}{Value}
\\
\hline
\hline
$M_Z $ & $ j_T^{had}$  & --0&77
\\
$M_Z $ & $ j_T^{lept}$ & --0&47
\\
$\Gamma_Z $ & $ r_T^{had}$ & 0&80
\\
$\Gamma_Z $ & $ r_T^{lept}$ & 0&62
\\
$r_T^{had} $ & $ r_T^{lept}$ & 0&78
\\
$j_T^{had} $ & $ j_T^{lept}$ & 0&49
\\  \hline
\end{tabular}

\vspace*{5mm}
\caption
{\it
Biggest correlations in the S-matrix fit
\label{table2}
}
\end{table}

To summarize, from both the strong experimental correlations in the
S-matrix fit  and the excellent 
agreement of the central values of fitted parameters in all fit scenarios one
may conclude that the different scenarios are highly compatible with each
other.

\section*{Acknowledgements}
I would like to thank Prof. Antonov and Dr. Christova, the organizers
of the Scientific Conference of the Theoretical Section of the 
Faculty of Physics on the Occasion
of the 25$^{th}$ Anniversary of the Bishop Konstantin 
Preslavsky University of Shoumen, Bulgaria, held at October 31, 1996, 
for the invitation to present a lecture and for the generous
hospitality. 
\\
I also would like to mention that I am grateful to Penka Christova for
the  long enjoyable and fruitful scientific cooperation.

\bigskip

An extended version of this write-up is DESY 97-218.

{ \small
\def\href#1#2{#2}
\bibliography{%
/home/phoenix/riemann/Bibliography/ca,%
/home/phoenix/riemann/Bibliography/radcorr,%
/home/phoenix/riemann/Bibliography/basics,%
}
\bibliographystyle{/home/phoenix/riemann/Bibliography/utphys_t}
} 

\end{document}